\documentclass[10pt, letterpaper]{article} %% two column, final layout

\usepackage{opex3}
\usepackage[draft]{hyperref}
\usepackage{amsmath}
\usepackage{setspace}
\usepackage[english]{babel}

\begin{document}

\title{Sum-frequency generation of 589~nm light with near-unit efficiency}

\author{Emmanuel Mimoun$^1$, Luigi De Sarlo$^1$, Jean-Jacques Zondy$^2$, Jean Dalibard$^1$, and Fabrice Gerbier$^1$}

\address{$^1$ Laboratoire Kastler Brossel, ENS, Universit{\'e} Pierre et Marie-Curie-Paris 6, \
CNRS, 24 rue Lhomond, 75005 Paris, France. \\
$^2$ Institut National
de M\'etrologie (LNE-INM-Cnam), 61 rue du Landy, 93210 La Plaine Saint
Denis, France. }

\email{emmanuel.mimoun@ens.fr}

\begin{abstract}
We report on a laser source at 589~nm  based on sum-frequency generation of two infrared laser at $1064{\rm~nm}$ and $1319{\rm~nm}$. Output power as high as 800~mW is achieved starting from 370~mW at 1319~nm and 770 mW at 1064~nm, corresponding to converting roughly 90\% of the 1319~nm photons entering the cavity. The power and frequency stability of this source are ideally suited for cooling and trapping of sodium atoms. 
\end{abstract}

\ocis{140. 7300, 190. 2620, 140. 3425, 020. 3320}

%\maketitle

%] %% activate for two-column option
\bibliographystyle{osajnl}

 %%%%%%%%%%%%%%%%

\section{Introduction}

Among all elements of the periodic table, Sodium historically played a key role in the development of atomic physics and spectroscopy,  and still retains a considerable importance for  fundamental research or applications such as artificial beacon stars, Laser-induced detection in the atmospheric range
(LIDAR)~\cite{fugate1991a}, and quantum degenerate gases. However, reaching the yellow resonance (''Sodium doublet'') near 589~nm requires to use dye lasers, which are expensive and difficult to maintain and operate. For this reason, many alternative methods based on non-linear freqency conversion of solid-state infrared lasers have been explored~\cite{jeys89a,moosmueller97a,vance98a,bienfang2003a,denman2004a,mildren2004a,feng2004a,janousek2005a,georgiev2006a,dawson2006a}. Second harmonic generation from a Raman fiber laser~\cite{mildren2004a,feng2004a,georgiev2006a}  and sum frequency generation (SFG) from two lasers around $938{\rm~nm}$ and $1583{\rm~nm}$~\cite{dawson2006a}, or around $\lambda_1=1064{\rm~nm}$ and $\lambda_2=1319{\rm~nm}$, have been demonstrated. This last solution seems particularly appealing since both wavelengths are accessible using YAG lasers. Such infrared sources in cavity-enhanced configurations have been demonstrated~\cite{jeys89a,vance98a,bienfang2003a,denman2004a,janousek2005a}. Up to $P_3=20~$ W output power at $\lambda_3=589{\rm~nm}$ has been reported~\cite{bienfang2003a,denman2004a}, based on two custom high-power ($P_1$=$20\ {\rm W}$ and $P_2$=$15\ {\rm W}$) infrared lasers. 

Here we report on the experimental realization of a 589~nm source with up to $P_3 \approx 800~{\rm mW}$ output power at $\lambda_3=589{\rm~nm}$, using moderate infrared powers from commercial laser sources. Our system operates in a highly efficient regime, where roughly 90\% of the photons of the weakest ($1319\ {\rm~nm}$) source effectively coupled into the cavity are converted. We show below that the resulting depletion of the $1319\ {\rm~nm}$ pump source strongly distorts the cavity fringe pattern. We have therefore designed and implemented an original fringe reshaping method to efficiently and robustly lock the lasers to the cavity.  Using the resulting $589{\rm~nm}$ laser source, a magneto-optical trap of sodium is obtained, confirming a linewidth of the laser below the natural linewidth ($10~$MHz) of the atomic transition. To our knowledge, this had only been achieved previously using dye lasers (see {\it e.g.} \cite{davis95a,streed2006a}).  

The paper is organized as follows. In Section~\ref{section2}, we present the main elements in our experimental setup. In Section~\ref{section3}, the theory of cavity-enhanced SFG is recalled, with particular emphasis on how to reach an optimal regime where almost all incoming photons from the weakest source are converted. In Section~\ref{section4}, we discuss experimental issues associated with large conversion efficiencies, and how to resolve them. Experimental results are presented in Section~\ref{section5}, and conclusions are exposed in Section~\ref{section6}.

%Although the range of optical wavelengths that can be reached with solid-state %lasers keeps broadening, building a cost-effective solution to produce a powerful single-frequency laser source in the yellow-orange ($550~-~600{\rm~nm}$) spectral range remains an open challenge. Non-solid options such as dye lasers or copper vapor lasers have been used for a long time, but they are expensive and difficult to maintain and operate. An alternative approach uses non-linear processes to generate visible light from solid-state infrared lasers\cite{jeys89a,moosmueller97a,vance98a,bienfang2003a,mildren2004a,feng2004a,janousek2005a,georgiev2006a,dawson2006a}. Second harmonic generation from a Raman fiber laser~\cite{mildren2004a,feng2004a,georgiev2006a}  and sum frequency generation (SFG) from two lasers around $938{\rm~nm}$ and $1583{\rm~nm}$~\cite{dawson2006a}, or around $\lambda_1=1064{\rm~nm}$ and $\lambda_2=1319{\rm~nm}$, have been demonstrated. The latter solution is particularly appealing, both wavelengths being accessible using YAG lasers. Such infrared sources have been used in both single-pass~\cite{moosmueller97a} (generating little power) and cavity-enhanced configurations~\cite{vance98a,bienfang2003a,janousek2005a,jeys89a}. However, to our knowledge, no such system has combined a narrow linewidth ($\lesssim 1~{\rm MHz}$), {\bf a near-unit conversion efficiency, and output and input laser powers both in the watt-level range}. 

\section{Experimental setup}\label{section2}

\begin{figure}
\centerline{\includegraphics{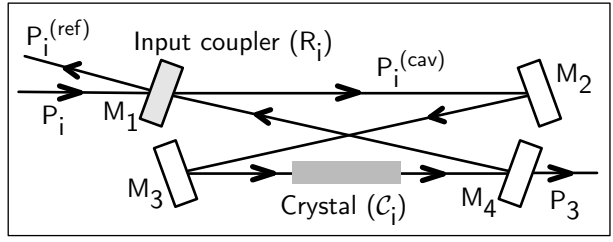}}
\caption{Setup for SFG in a doubly resonant bow-tie cavity formed by four mirrors $M_{1,..,4}$. For the pump lasers $i$ ($i$~=~$1,2$), we note $P_i^{\rm (cav)}$ the intra-cavity power, $P_i$ the incident power, $P_i^{\rm (ref)}$ the reflected power, $C_i$ the power fraction transmitted through the crystal, ad $R_i$ the reflectivity of the input coupler. $P_3$ is the output power produced at $\lambda_3$. }
\label{fig:schema_cavity}
\end{figure}

Our system is described schematically in Fig.~\ref{fig:schema_cavity}. The lasers operating around $\lambda_1=1064{\rm~nm}$ and $\lambda_2=1319{\rm~nm}$ respectively are coupled into a bow-tie cavity, which is resonant for both of them and transparent for the output laser at $\lambda_3=589~$nm.  For these lasers we use commercial sources (manufactured by Innolight GmbH, Germany) delivering $1~{\rm W}$ at $1064{\rm~nm}$ and  $500~{\rm mW}$ at $1319~{\rm~nm}$.
The useful powers effectively coupled into the cavity are lower, $P_1 \approx 770~{\rm mW}$ at $1064{\rm~nm}$ and $P_2 \approx 370~{\rm mW}$ at $1319{\rm~nm}$, due to losses through the optical path and a measured 85\% coupling efficiency into the fundamental mode of the cavity. Inside the cavity we place a periodically poled KTiOPO$_4$ (ppKTP) crystal (manufactured by KTH, Sweden), with a poling period $\approx~12.7~\mu$m. The choice of ppKTP was motivated by its relatively high non-linear coefficient, by its tolerance to large input powers, and by its negligible absorption in the visible range. The crystal is enclosed in a copper mount which is temperature-regulated within a few 10~mK. We found an optimal single-pass operating temperature around 50$^\circ$C, with single-pass efficiency $\alpha=P_3/P_1 P_2\approx0.022~$W/W$^2$, in reasonable agreement with the previously measured values ~(e.g.~\cite{janousek2005a}).

\section{Sum-frequency generation in cavity-enhanced configuration}\label{section3}

In this Section, we first recall the theory of SFG and discuss how it can be used to optimize the desired $589$~nm output power. In our theoretical model, we assume that only a small fraction of the intra-cavity powers is consumed in the SFG process. In this weakly-depleted pumps approximation, the output power $P_3$ at wavelength $\lambda_3$  is simply given by~\cite{boyd2003a,boyd1968a,zondy1997a}:

\begin{equation}
	\label{eq:wdp}
	P_{3}=\alpha\ P_{1}^{\rm (cav)} P_{2}^{\rm (cav)}.
\end{equation} 
where $P_{i}^{\rm (cav)}$ is the intra-cavity power at wavelength $\lambda_i$. On resonance, the intra-cavity power $P_{i}^{\rm (cav)}$ is linked to the incident power $P_i$ by~\cite{siegman1986a}
\begin{eqnarray}
\label{eq:celoss}
\frac{P_{i}^{\rm (cav)}}{P_i}=\frac{1-R_i-L_i}{\left(1-\sqrt{R_i(1-\delta_i) C_i}\right)^2}.
\end{eqnarray}
Here, $R_{i}$ and $L_i$ denote reflection and loss coefficients of the input coupler $M_1$ for lasers $i\ (i=1,2)$, and $\delta_i$ denote the total passive fractional loss per cavity round trip. The  coefficient $C_i$ is the ratio between the photon flux before and after the crystal, and accounts for the depletion of laser $i$ due to SFG. Neglecting absorption in the crystal, energy conservation in the conversion process implies $C_i=1-\lambda_3 P_3/\lambda_i P_i^{\rm (cav)}$. As long as $C_i\simeq 1$, the weakly depleted pumps approximation is valid. 
%\begin{eqnarray}
%\label{eq:ce}
%\frac{P_{i}^{\rm (cav)}}{P_i}=\frac{1-R_i}{\left(1-\sqrt{R_i {\cal C}_i}\right)^2},
%\end{eqnarray}

Since SFG removes one photon from each infrared pump for each output photon, the efficiency of SFG is ultimately limited by the weakest pump $2$. Neglecting losses, the maximal output flux  at $\lambda_3$ equals the input flux at $\lambda_2$, corresponding to $P_3^{\rm(max)}=(\lambda_2/\lambda_3)P_2$. In the following, we take the ratio 
\begin{equation}
\eta= P_3/P_3^{\rm(max)}
\end{equation} 
 as a figure-of-merit for the conversion efficiency. This ratio compares the outcoming flux at $\lambda_3$ to the limiting incoming flux effectively coupled into the cavity at $\lambda_2$. With this definition, the conversion efficiencies reported in \cite{jeys89a,moosmueller97a,vance98a,bienfang2003a,mildren2004a,feng2004a,janousek2005a,georgiev2006a,dawson2006a} are generally around 50~\% (up to 60\% for \cite{bienfang2003a}).

\begin{figure}[t]
\centerline{\includegraphics[width=12cm]{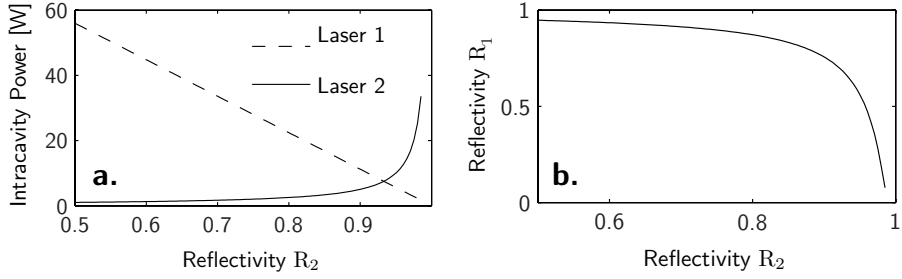}}                                             
 \caption{Lossless model : {\bf a.} Intracavity powers for laser 1 and 2 ($\lambda_1=1064{\rm~nm}$ and $\lambda_2=1319{\rm~nm}$) required to reach total conversion, plotted against the reflectivity $R_2$ of the input coupler; {\bf b.} Reflectivity $R_1$ of the input coupler plotted againt $R_2$ : Each couple ($R_1$,$R_2$) on this curve ensures total conversion.}
 \label{fig:lossless}
\end{figure}

Let us first consider an ideal situation with no passive losses in the cavity ($L_i$=0 and  $\delta_i$= $0$). In this case complete conversion of the weakest pump $2$ is possible ($P_3=P_3^{\rm(max)}$). From Eq.~(\ref{eq:celoss}), we find that this occurs when $C_2^{\rm opt}=R_2$, corresponding to the impedance matching between the power transmitted through the input coupler and that consumed in the crystal due to conversion~\cite{kaneda1997a}. The corresponding cavity enhancement factor is $P_2^{\rm (cav)}/P_2=(1-R_2)^{-1}$. Eq.~(\ref{eq:wdp}) then fixes the corresponding value for $P_1^{\rm (cav)}=(1-R_2) \lambda_2/\alpha \lambda_3$, which is achieved for a reflectivity $R_1$ that can be  computed from Eq.~(\ref{eq:celoss}). Without losses, one can always find a solution, implying that total conversion can be obtained in the lossless case for any value of $R_2$ (see Fig.~\ref{fig:lossless}({\bf a-b})). Consequently, the power reflected at mirror $M_1$
\begin{equation}
\label{eq:cr}
\frac{P_{2}^{\rm (ref)}}{P_2}=\frac{(\sqrt{C_2}-\sqrt{R_2})^2}{\left(1-\sqrt{R_2 C_2}\right)^2}. 
\end{equation} 
vanishes under these optimum conditions: all the incoming photons are converted in the crystal and no photon at $\lambda_2$ comes out of the cavity. This can be interpreted as a destructive interference on $M_1$ between the reflected field and the transmitted one. 
Note that the assumption of complete conversion does not contradict the weakly-depleted pumps approximation underlying the calculation. Indeed, because of the cavity power enhancement, the incident flux is a small fraction of the power inside the cavity.

\begin{figure}[t]
\centerline{\includegraphics{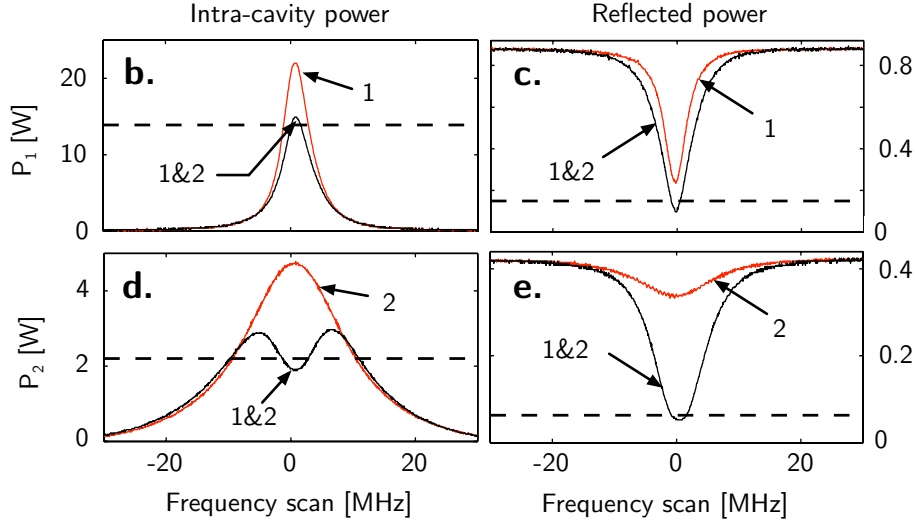}}
 \caption{{\bf b-d.}~Intra-cavity powers for lasers $1$~({\bf b}) and $2$~({\bf d}) ($\lambda_1=1064\ {\rm~nm}$, $\lambda_2=1319\ {\rm~nm}$), plotted against the cavity resonance frequency tuned with a piezoelectric transducer; {\bf c-e.}~Powers reflected out of the cavity for lasers $1$~({\bf c}) and $2$~({\bf e}). The dashed lines show the predictions from the model described in the text, including both conversion and passive losses.}  
\label{fig:scan_cavity}
\end{figure}

This lossless model is already sufficient to interpret qualitatively our experimental findings. In Fig.~\ref{fig:scan_cavity} we show the transmitted ({\bf b},{\bf d}) and reflected ({\bf c},{\bf e}) power while scanning the cavity length around the position where both lasers are simultaneously resonant. When only one laser is present (either $1$ or $2$) we observe the expected Lorentzian profile. However when both lasers are present a pronounced dip appears in the resonance profile for laser $2$, corresponding to efficient conversion into $589\ {\rm~nm}$ photons. A dramatic decrease of the reflected power is simultaneously observed, corresponding to the destructive interference previously mentioned. 

%\begin{table}
%\caption{Cavity and pump lasers parameters used in the SFG source. }
%\label{tab:param}
%\begin{tabular}{r|ccccc}
%&$R_i$&$L_i$&$P_i [W]$&$\kappa_i$&${\cal R}$\\ \cline{1-6}
%Laser 1 &0. 92&0. 012&0. 9&0. 85&\multirow{2}*{0. 98}\\
%Laser 2 &0. 74&0. 01&0. 44&0. 85&
%\end{tabular}
%\end{table}

\begin{figure}[t]
\centerline{\includegraphics{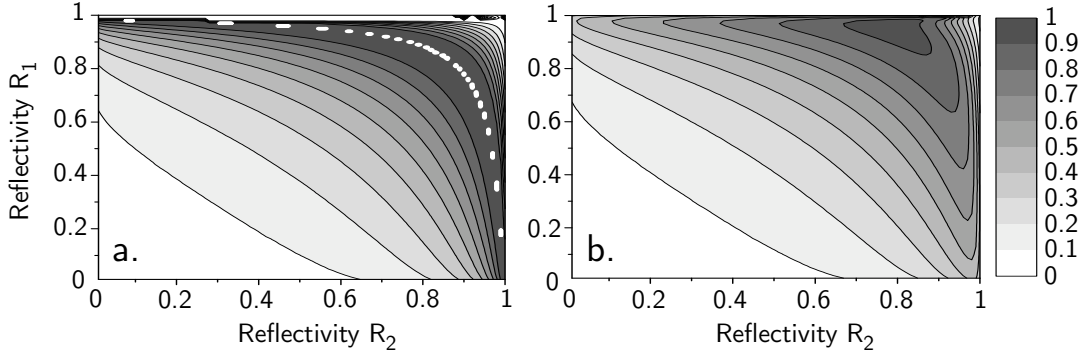}}                                             
 \caption{Contour map of the efficiency $\eta$ of the conversion process as a function of the reflectivities $R_1$ and $R_2$ of the input coupler $M_1$ for laser 1 and 2 ($\lambda_1=1064{\rm~nm}$ and $\lambda_2=1319{\rm~nm}$). {\bf a.}~Lossless case. The dashed line (same as in Fig.~\ref{fig:lossless}{\bf b}) corresponds to total conversion ($\eta=1$). {\bf b.}~Passive losses are taken into account using $(\delta_1=2.4\%,\delta_2=1.6\%)$. While $\eta=1$ cannot be obtained anymore experimentally, efficiencies higher than 90\% can still be reached.}
 \label{fig:r1vsr2_contour}
\end{figure}

The more realistic model including passive losses in Eq.~(\ref{eq:celoss}) 
can be solved numerically for a given single-pass efficiency and cavity parameters. The passive loss coefficients ($\delta_1=2.4\%,\delta_2=1.6\%$) have been determined by injecting only one laser at a time in the cavity, and comparing the measured transmitted and reflected powers to the theoretically expected values. The calculated conversion efficiency in this situation is shown in Fig.~\ref{fig:r1vsr2_contour}{\bf b}, as a function of the input coupler reflectivities. While total conversion cannot be achieved as in the lossless case (Fig.~\ref{fig:r1vsr2_contour}{\bf a}), a locus of points with a maximal efficiency close to 1 can still be identified. Among them, a sensible choice is to select $R_1$ and $R_2$ close to each other to minimize the total intra-cavity power and  thermal effects in the crystal. Note also that  the optimum is quite loose and the reflectivities relatively low, making the input coupler tolerant to small fabrication imperfections. 

\section{Locking scheme in the regime of large conversion}\label{section4}

Experimentally, working in a regime with such large conversion efficiencies leads to serious stability problems for a conventional locking system. The key to efficient SFG operation is to ensure that the cavity is simultaneously resonant with both IR lasers at all times. This is usually enforced by two servo-loops maximizing the intra-cavity powers independently.  The locking scheme is as follows in our experiment. The 1319~nm laser is used as a
master laser onto which the cavity length is locked using an
integrating servo-loop. Then, the 1064 laser is locked
onto the cavity, and therefore on the master laser, ensuring
stable operation of the ensemble. In our experiment each servo loop uses an error signal generated from the power leaking through one of the cavity mirror by a modulation/demodulation technique. However what follows would still be valid for other locking techniques.

In the regime of large conversion, methods relying on such a master/slave scheme fail due to the above mentioned dip in the transmission of laser 2 (see Fig.~\ref{fig:scan_cavity}{\bf d}). Indeed, the cavity servo cannot distinguish this power reduction from that caused by an external perturbation, and actually works against keeping both lasers on resonance simultaneously. To circumvent this problem, we have designed an original analog processing of our error signals. Instead of the ''bare'' error signal produced by laser $2$, the cavity lock uses a linear combination of this signal and of the output at $\lambda_3$. The combination is done electronically, with weights chosen empirically to restore a lineshape with a single maximum and optimize the slope around the lock-point. This fringe reshaping method works for any level of conversion, and allows stable operation of the laser on a day timescale, even at the highest efficiencies. Finally, choosing the 589~nm output as the error signal for the second servo-loop locking laser $1$ to the cavity ensures that the system locks to the maximal converted power. This fringe reshaping method is protected by French patent INPI 0803153 (international patent pending), and further details will be given in a future publication~\cite{mimoun2008b}. 

\section{Experimental results}\label{section5}

\begin{figure}[t]
\centerline{\includegraphics{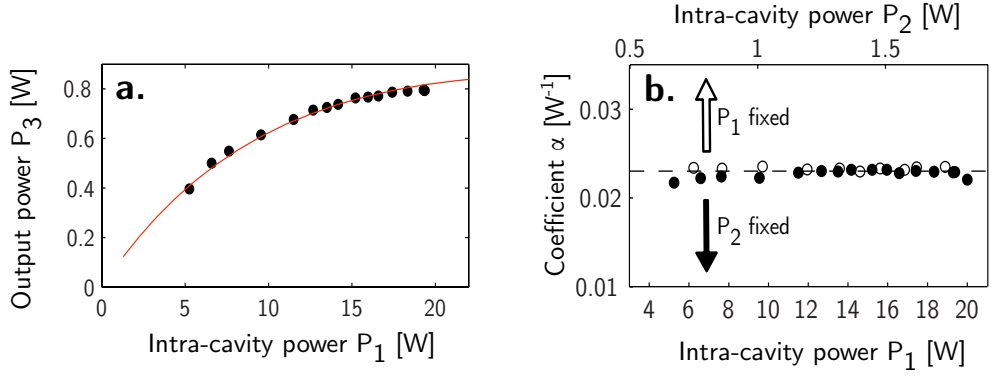}}
 \caption{{\bf a.}~Output power at 589~mn plotted against intra-cavity power at 1064~\rm{nm} (varied by changing the incoming power into the cavity). The solid line is the result of the numerical calculations as described in text. {\bf b.}~Conversion coefficient $\alpha$ (see Eq.~(\ref{eq:wdp})), varying the incoming power of one pump laser while leaving the other fixed. The conversion coefficient is constant and equal to that measured in the single pass configuration (dashed line), irrespective of laser power.}    
\label{fig:conv_coeff_plus}
\end{figure}

We have performed a systematic study of the dependance of the output power on the infrared pump powers. Such a measurement is shown in Fig.~\ref{fig:conv_coeff_plus}a. The model including passive losses ($\delta_1=2.4\%,\delta_2=1.6\%$) compares favorably to our experimental findings. Both the calculated steady state values (dashed lines in Fig.~\ref{fig:scan_cavity}({\bf b}-{\bf e})) and the variation of output power $P_3$ with pump power $P_1$  (shown in  Fig.~\ref{fig:conv_coeff_plus}{\bf a}) are well-reproduced by the model. Furthermore, Fig.~\ref{fig:conv_coeff_plus}{\bf b} shows no variation of the coefficient $\alpha$ deduced from Eq.~(\ref{eq:wdp}) even at the highest powers, thus validating the weakly depleted pump approximation. This also rules out additional effects (such as thermal effects related to absorption of the infrared beams) which would reduce $\alpha$ at higher powers. Overall, we find that the model gives a reliable description of the SFG process for our experimental configuration, and allows one to optimize the parameters of the cavity to ensure maximum efficiency. A key parameter to achieve $\eta\simeq 0.9$ is the choice of a highly nonlinear crystal, resulting in a nonlinear loss $\alpha P_1^{\rm (cav)}$ exceeding by far the roundtrip passive loss. With our set of parameters, we were able to reach output powers as high as $P_3=800~{\rm mW}$ or $\eta \approx 90\%$. This implies that our apparatus works very close to the theoretical ideal limit studied in the first part of the paper. 

\begin{figure*}
\centerline{\includegraphics{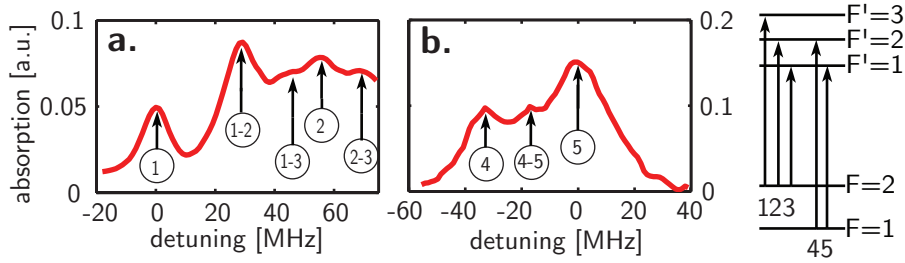}}
\caption{Saturated absorption signal (solid line) while scanning the
frequency of the 1319nm laser, thus the one of the 589nm one. The D2
transitions corresponding to atoms in the ground F=2 ({\bf a}) and F=1 ({\bf b})
electronic states are represented. $(i-j)$ represents the level crossing line between transitions $i$ and $j$.} \label{fig:absorption}
\end{figure*}

In our experiment, the laser is locked on the Sodium $D_2$ resonance line by reacting on the frequency of laser 2 using an error signal obtained from saturated absorption spectroscopy, as shown in Fig.~\ref{fig:absorption}. Using this laser source, we have obtained a magneto-optical trap containing roughly $10^7$ atoms in a ultra-high vacuum cell. Repumping light could be derived from the same source as the trapping laser itself by using a high-frequency ($1.7$~GHz) acousto-optic modulator (manufactured by Brimrose Corporation of America). This confirms the viability of our approach for demanding applications such as laser cooling and trapping, or high resolution spectroscopy.

\section{Conclusion}\label{section6}

In conclusion, we presented an efficient all-solid state laser source based on SFG at $589\ {\rm~nm}$  with an output power of $800~{\rm mW}$. The source acts as a wavelength converter for the weakest source,  ensuring a conversion efficiency around 90\% while keeping input powers in the watt-level range. Such a setup can be used to produce other wavelengths in the visible range, provided the existence of input lasers at the right wavelengths. This provides a cost-effective solution for atomic physics experiment, free from the drawbacks of dye lasers. 

\section*{Acknowledgements}
We thank Pierre Lemonde, Antoine Browaeys, Wolfgang Ketterle, Aviv Keshet and Volker Leonhardt for discussions. This work was supported by ANR (Gascor contract), IFRAF, DARPA (OLE project), and EU (MIDAS network). LdS acknowledges financial support from  IFRAF. JJZ acknowledges funding from Laboratoire National de M\'etrologie et d'Essais (LNE).

% section exp (end)

\end{document}